\documentclass[english]{sig-alternate}
\usepackage[T1]{fontenc}
\usepackage[latin9]{inputenc}
\usepackage{listings}
\usepackage{graphicx}
\usepackage[numbers]{natbib}

\makeatletter
 \newtheorem{problem}{Problem}

\makeatother

\usepackage{babel}
\begin{document}

\title{Efficient programs of NPC problems should be length upper-bounded,
and a thought experiment to search for them by machine enumeration}

\author{\alignauthor YuQian Zhou\titlenote{This work is supported by personal funding after I quit from my previous company and founded my own startup.}\\                \email{zhou@tuuso.com}}
\maketitle
\begin{abstract}
This paper proposes a thought experiment to search for efficient bounded
algorithms of NPC problems by machine enumeration. The key contributions
are:
\begin{itemize}
\item On Universal Turing Machines, a program's time complexity should be
characterized as: execution time(n) = loading time(n) + running time(n).{\small \par}
\item Introduces the concept of bounded algorithms; proposes a comparison
based criterion to decide if a bounded algorithm is inefficient; and
establishes the length upper bound of efficient bounded programs.{\small \par}
\item Introduces the growth rate characteristic function to evaluate program
complexity, which is more easily machine checkable based on observations.{\small \par}
\item Raises the theoretical question: if there exists any bounded algorithm
with polynomial execution time for NPC problems.{\small \par}
\end{itemize}
\end{abstract}
\markboth{YuQian Zhou}{Efficient programs of NPC problems are length upper-bounded}

\section{Motivation}

It has been for decades since the P ?= NP question was introduced
by \citep{DBLP:conf/stoc/Cook71}, but people still have not found
a polynomial algorithm to any of the NPC problems. This leads many
researchers to doubt if such algorithms exist at all. One way to either
confirm or dismiss the doubt is to exhaustively search for such algorithms.
But this is impossible because in theory there are \emph{infinite}
number of programs, since usually we do not limit a program's length.

However, let us consider a NPC problem with a particular input size
$n$. Suppose one of its solution program's length is in the order
of exponential of $n$, in practice on a general-purpose computer
(UTM), we will not consider such program \emph{efficient}, since its
loading time alone will take exponential time, regardless of the program's
running time complexity. And for a specialized computer (TM), exponential
length means the machine is too expensive to build, either from physical
material, or virtual ones such as digital bits. This means there should
a upper bound of the length of the program (with respect to the input
size) that we will consider efficient. Given such program length upper
bound, then the total number of programs is \emph{finite}, so we can
enumerate them, and check if there exists an efficient program for
a NPC problem. This is the main intuition that motivated this paper.

This paper is organized as follows. In section 2, we discuss program
complexity on Universal Turing Machines; in 2.1, we introduce the
concept of bounded algorithms; in 2.2, we propose a comparison based
criterion to decide if a bounded algorithm is inefficient; and in
2.3, we establish the length upper bound of efficient bounded algorithms.
In section 3, we introduce a new way to evaluate program complexity,
which is more easily machine checkable based on observations. In section
4, we propose a thought experiment to search for efficient bounded
algorithms of NPC problems by machine enumeration. In section 5, we
raise the question whether there exists bounded algorithm with polynomial
execution time for NPC problems, and discuss some possible implementation
issues with the thought experiment.

In this paper, when we talk about time complexity, it always means
worst time complexity; and we sometimes use the term ``algorithm''
and ``program'' (that implements the algorithm) interchangeably.

\section{Program complexity on Universal Turing Machines}

Traditionally in theoretic computer science literature, an algorithm's
time complexity quantifies the algorithm's \emph{running} time. It
does not care about the algorithm's\emph{ loading} time. The reason
is that when discussing time complexity, the computation model used
is Turing Machine (TM, either deterministic or non-deterministic),
which computes a fixed partial computable function (the algorithm).
The machine description is pre-built, therefore there is no loading
time, or we treat it always as $0$.

However, most of the computers people use today are the general-purpose
computers. They are modeled on Universal Turing Machines (UTM), which
first reads (loads) an arbitrary stored-program, i.e. the description
of a TM (algorithm), and then simulates it on arbitrary input. Therefore:
\begin{description}
\item [{Axiom}] On a UTM, for input size $n$, let the program's total
execution time be $E(n)$, loading time be $L(n)$, and running time
be $R(n)$, then
\[
E(n)=L(n)+R(n)
\]

\item [{Theorem}] On UTM, if there exists an algorithm that solves a NPC
problem in polynomial execution time, then both its loading time and
running time should be polynomial.
\end{description}
From now on, when we discuss a program's \emph{execution} time complexity
on a UTM, we also need to investigate the program's loading time.
It is natural to assume that, for theoretical UTMs (and practical
general-purpose computers):
\begin{description}
\item [{Axiom}] A program's loading time is linear to the program's length.
\end{description}
In computer science textbook, most algorithm's length is constant,
and can handle all input size $[0,+\infty)$. As the input size $n$
increases big enough, the running time $R(n)$ (normally a monotonically
increasing function) will dominate the equation, so $L(n)$ can be
ignored.

However on real world computers, programs can only deal with problems
with input of finite size, because of either space (memory or disk)
or time constraints. When the $n$ is not big enough, or when an algorithm's
length bears some relationship with the input size $n$, then $L(n)$
cannot be ignored.

As an example, in the following we will construct a program called
PRECOMPUTE: it is a polynomial running time algorithm for the 3-coloring
problem of undirected n-nodes graphs (a NPC problem) with input size
$[0,n]$ for some fixed $n$, but its loading time is exponential
to $n$. First we define program SIMPLE which will be used later to
construct PRECOMPUTE.
\begin{description}
\item [{Example:~SIMPLE}] Generate all the possible combinations of nodes
3-coloring schemes, then try them one by one to see if there is any
conflicts, i.e. two adjacent nodes have the same color:

\begin{lstlisting}
# input: graph as the set of nodes & edges
def SIMPLE(g={nodes, edges}):
  colorings = generate_all_combinations(nodes, [R,G,B])
  for c in colorings:
    if no_conflict(edges, c):
      return true
  return false
\end{lstlisting}

\end{description}
The running time of SIMPLE is exponential (i.e. $O(3^{n})$, where
$n$ is the number of nodes). Now let us construct PRECOMPUTE:
\begin{description}
\item [{Example:~PRECOMPUTE}] Label each graph node with a unique number:
$\{0,1,...,n-1\}$, denote the edge between two nodes $x$ and $y$
as $e_{(x,y)}$, where $x<y$. There are $|E|=\frac{n\times(n-1)}{2}$
possible edges between any two nodes, and there are $|G|=2^{|E|}$
possible graphs with n nodes (we do not consider graph iso-morphism).

Label each possible edge with a unique prime number, i.e. $h(e_{(x,y)})=p_{i}$
where $i\in[0,|E|-1]$ and $p_{i}$ is the i-th prime number. Now
each of the possible graph $g\in G$ can be uniquely labeled with
a number by taking the products of all its edge labels: $h(g)=\prod_{e\in g}h(e)$.
In computer science terms, the $h$ we have just defined is the hash
function of a graph.

Now let us construct the program PRECOMPUTE: for each of the possible
graph $g\in G$, using SIMPLE to calculate if it can be 3-colored,
and record the result as a pair $(h(g),r)$ into a hashtable, where
$r$ is $true$ or $false$ depending on whether $g$ can be 3-colored
or not. Output the hashtable as the data segment of PRECOMPUTE.

The code segment of PRECOMPUTE is: for input graph $g$,
\begin{lstlisting}
# input: graph as the set of nodes & edges
def PRECOMPUTE(g={nodes, edges}):
  hashtable = [...(pre-computed static data)...]
  if nodes.length > n:  # check input size
    return undefined
  else
    key = h(g)
    r = hashtable[key]
    return r
\end{lstlisting}

Line 7: calculate $key=h(g)$, which takes at most $O(|edges(g)|)$,
i.e. time linear to the number of input edges

Line 8: look up the result ($true$ or $false$) from the hashtable
using $key$, which takes $O(1)$ time

So the running time of PRECOMPUTE is clearly polynomial, while its
loading time is exponential to $|E|$.

\end{description}

\subsection{Bounded algorithm}

The program PRECOMPUTE we have just constructed is different from
usual programs in that it can only handle input upto a fixed size.
Now let us introduce the concept of bounded algorithm.
\begin{description}
\item [{Definition:~bounded~algorithm}] given a number $n$, if an algorithm
$A$ returns $correct\, result$ for any input of size $\leq n$;
and returns either $correct\, result$ or $undefined$ for input size
$>n$, then we say $A$ is an \textit{$n$-bounded} algorithm; if
algorithm $A$ always (theoretically) returns $correct\, result$
for all input size $[0,+\infty)$, then we say algorithm $A$ is \textit{unbounded}.
\end{description}
The set of unbounded algorithms is clearly a proper subset of the
set of bounded algorithms. In the real world computers can only work
on problems of finite size, bounded algorithms will actually give
programmers more design and implementation choices.

Most algorithms in computer science literature are unbounded, e.g.
Euclid's GCD algorithm, and any sorting algorithms; and their length
are constant (with respect to the input size $n$). So even on UTM,
an unbounded algorithm's loading time can be ignored as $n$ becomes
significantly large enough:
\[
E(n)=R(n)
\]

For bounded algorithm, the algorithm's length can be related to the
input size, thus play an important role in the execution time complexity,
just as we have shown in PRECOMPUTE.
\begin{description}
\item [{Example:~Search~Engine}] It is also tempting to load a program
once, and run it multiple times, so the execution time equation becomes:
\[
E(n)=L(n)+m\times R(n)
\]
And when $m\rightarrow\infty$, $L(n)$ can be ignored. E.g. in such
case, the PRECOMPUTE we just constructed becomes a search engine for
the graph 3-coloring problem, whose running time (query response time)
as appeared to the user is polynomial.
\end{description}
Since we have constructed an algorithm PRECOMPUTE with this property,
we will not discuss such use case any more. In the remaining of this
paper, we will only consider bounded algorithms with $m=1$.

\subsection{A comparison based criterion to decide if a bounded algorithm is
inefficient}

In the program PRECOMPUTE, by the way it is constructed, we know that
its length is exponential to the input size $L(n)=O(2^{|E|})$, so
we can decide it is inefficient. Now suppose we do not know how this
program is constructed (e.g. imaging it is from an oracle), and we
lack necessary tools to analyze the relationship between its length
and the input size. Then, what criterion we should use to decide that
this program is\textit{ inefficient} for input size $n$?

In the following, we introduce a comparison based criterion:
\begin{description}
\item [{Definition:~UTM-inefficient}] For a particular problem with input
size $n$, on a fixed UTM, let $known\_inefficients$ be the finite
set of all the bounded programs that human know so far (by some other
means, e.g. source code analysis) are inefficient.\\
Let $wet(prog(n))$ be the worst execution time of program $prog$
with input size $n$, and we denote the minimum of the worst-case
execution time of all those programs in $known\_inefficients$ as
$minwet(n)$, i.e. 
\begin{eqnarray*}
 &  & minwet(known\_inefficients,n)\\
 &  & =min_{prog\in known\_inefficients}(wet(prog(n)))
\end{eqnarray*}
Let $A$ be a $n$-bounded algorithm, if $A$'s execution time is
longer than any known inefficient algorithm, i.e. $E(A,n)\geq minwet(known\_inefficients,n)$
then $A$ is called \emph{UTM-inefficient} for the given input size
$n$.
\end{description}
For example, initially we can add SIMPLE, and PRECOMPUTE to the knowledge
base $known\_inefficients$:
\begin{itemize}
\item SIMPLE, length complexity $O(1)$, running time complexity $O(3^{|nodes|})$.
\item PRECOMPUTE, length complexity $O(2^{|E|})$, running time complexity
$O(|edges|)$.
\end{itemize}
Note, as the human knowledge ($known\_inefficients$) increases, $minwet(n)$
will decrease.

\subsection{Length upper bound of efficient bounded programs}
\begin{description}
\item [{Corollary}] If a $n$-bounded algorithm's length $\geq minwet(n)$,
then it is \emph{UTM-inefficient}.\end{description}
\begin{proof}
$E=L+R$, and $L\geq minwet(n)$, therefore $E\geq minwet(n)$.
\end{proof}
Thus $minwet(n)$ is an upper bound of the length of efficient bounded
programs.

\section{A computable property of program complexity}

Let us exam the execution time complexity on UTM again:
\[
E(n)=L(n)+R(n)
\]
We have established the length upper bound of efficient bounded programs,
before we can start to enumerate programs on a UTM, and search for
efficient bounded programs for NPC problems, there is one more issue:
it will be better to have a method to evaluate an algorithm's running
time that is machine checkable.

Traditionally, the running time complexity of an algorithm is analyzed
by human. We take the algorithm's description (e.g. source code),
and use human knowledge and skills to establish a mathematical model
and formulate a limiting function of the program's running time with
respect to its input size. However this step cannot be easily formalized
and automated by a computer program. In the this section we will introduce
a method to evaluate program complexity that is more machine checkable
based on observations.
\begin{description}
\item [{Definition:~growth~rate~characteristic~function}] let $f(n)$
be the limiting function of the complexity of an algorithm with input
size $n$ in the big-O notation, i.e. $T(n)=O(f(n))$, for $n>1$
we define
\[
g(f(n))=\log_{n}f(n)
\]
as the \emph{growth rate characteristic function} of the algorithm.
\end{description}
Note, it does not matter whether it is time complexity $T(n)$, space
complexity $S(n)$, or any other kind of program complexities, the
following discussion apply to all of them.

Let us consider two important limiting functions in algorithm complexity
analysis: polynomial and exponential functions, let $k>0$ be constant:
\begin{itemize}
\item For polynomial complexity $T(n)=O(n^{k})$, $g(f(n))=k$
\item For exponential complexity $T(n)=O(2^{n^{k}})$, $g(f(n))=n^{k}\log_{n}2$
\end{itemize}

\subsection{Apply $g$ on observations}

Given a program, we can record its actual running steps corresponding
to a series of input size $[n_{0},n_{1},...n_{i}]$ as our observations.
We will study $g$'s properties on these observations. Let $ob(n)$
be the actual observed steps for the algorithm performed on input
of size $n$:
\begin{itemize}
\item For polynomial complexity $T(n)=O(n^{k})$: by the definition of big-O
notation, there exists constant $M>0$, such that $ob(n)\leq M\times n^{k}$,
where $n\geq n_{0}$ for some constant $n_{0}>1$, then
\begin{eqnarray*}
g(ob(n)) & = & \log_{n}ob(n)\\
 & \leq & \log_{n}M\times n^{k}\\
 & = & \log_{n}M+\log_{n}n^{k}\\
 & = & \log_{n}M+k
\end{eqnarray*}
and
\[
\lim_{n\to+\infty}g(ob(n))\leq\lim_{n\to+\infty}(\log_{n}M+k)=k
\]
Summary: the upper bound of $g(ob(n))$ has limit $k$; and it is
\emph{monotonic decreasing} with max value $(\log_{2}M+k)$ if $M>1$,
or \emph{monotonic increasing} with min value $(\log_{2}M+k)$ if
$M<1$.
\item For exponential complexity $T(n)=O(2^{n^{k}})$:
\begin{eqnarray*}
g(ob(n)) & = & \log_{n}ob(n)\\
 & \leq & \log_{n}M\times2^{n^{k}}\\
 & = & \log_{n}M+\log_{n}2^{n^{k}}\\
 & = & \log_{n}M+n^{k}\log_{n}2
\end{eqnarray*}
and
\begin{eqnarray*}
\lim_{n\to+\infty}g(ob(n)) & \leq & \lim_{n\to+\infty}(\log_{n}M+n^{k}\log_{n}2)\\
 & =\\
 &  & \lim_{n\to+\infty}n^{k}\log_{n}2\\
 & =\\
 &  & \lim_{n\to+\infty}\frac{n^{k}\ln2}{\ln n}\\
 & = & (\mathrm{L'H\hat{o}pital's\, rule})\\
 &  & \lim_{n\to+\infty}\frac{kn^{k-1}\ln2}{1/n}\\
 & =\\
 &  & \lim_{n\to+\infty}kn^{k}\ln2\\
 & = & +\infty
\end{eqnarray*}
Summary: the upper bound of $g(ob(n))$ has limit $+\infty$; and
it is\emph{ monotonic increasing} after sufficient large $n$%
\footnote{The proof is left as an exercise to the interested readers.%
}.\end{itemize}
\begin{description}
\item [{Example}] The following two figures illustrate the upper bound
function curves for polynomial and exponential complexity:\\
\includegraphics[scale=0.33]{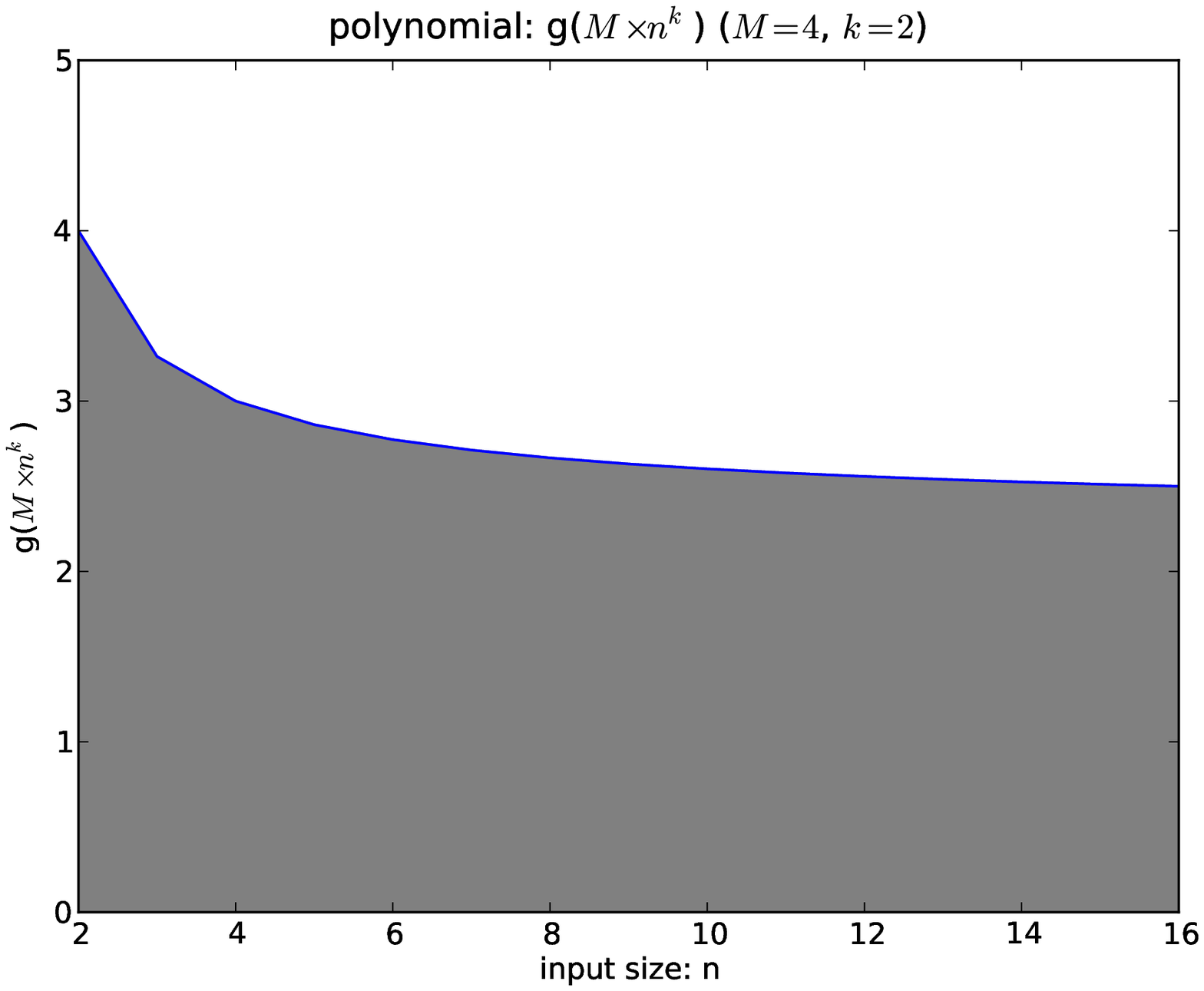}\\
\includegraphics[scale=0.33]{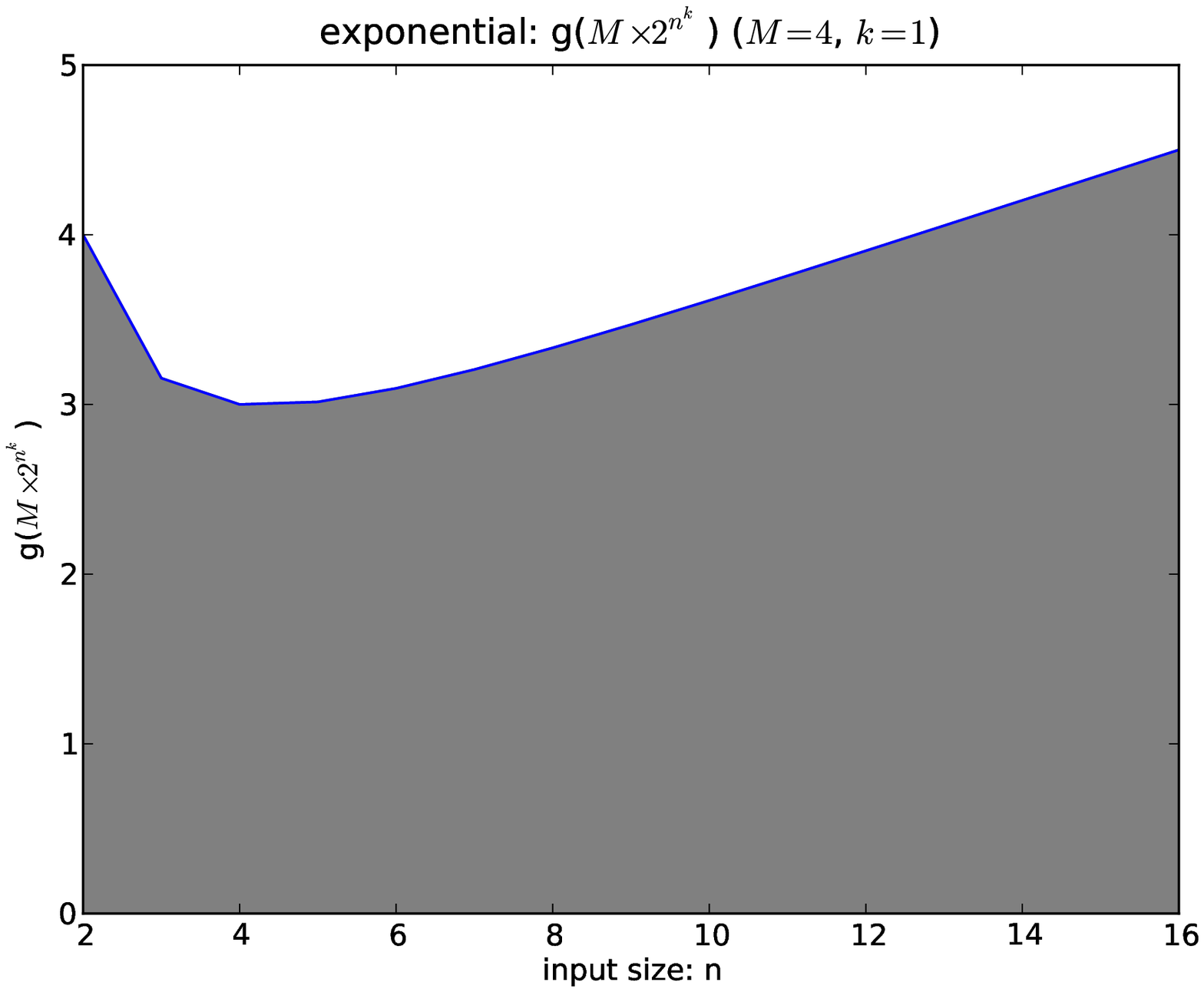}
\end{description}
Note: what we just discussed is the bounding function's property of
an algorithm, which is different from the actual observations. For
example, the $g$ of actual observations of a polynomial algorithm
can be oscillating, but still being bounded, e.g. a program that blankly
loops for $n^{2+cos(n\pi)}$ steps for input size of $n$.

\subsection{$g_{n_{e}}^{u}$ on finite observations}

Since we can only make finite observations, any algorithm's $g$ will
always be bounded by some value. For example both the algorithms in
the previous example are bounded by $g=5$ for observations on input
of size $[2-16]$. Because of the ``sufficient large $n$'' assumption,
we are more interested in the ending point metric. Let us introduce
a notation $g_{n_{e}}^{u(n_{e})}$ where $n_{e}$ is the ending observation
point $n$, and $u$ is the max value of $g(n)$ for all $n\in[2,n_{e}]$,
for simplification we use its ceiling integer value, i.e.
\[
u(n_{e})=ceiling(max_{n\in[2,n_{e}]}(g(n)))
\]
So in the previous example, the polynomial is a $g_{16}^{4}$ algorithm,
while the exponential is a $g_{16}^{5}$ algorithm.
\begin{description}
\item [{Algorithm~efficiency~evaluation~method:}] Given an algorithm
$A$, if for all sufficiently large observation points $n<m$, $u(m)\leq u(n)$,
then $A$ is a possible polynomial algorithm.
\end{description}

\section{A thought experiment to search for efficient \emph{$n$}-bounded
algorithms of NPC problems by machine enumeration}

In the previous sections, we have established the length upper bound
$UB$ of efficient bounded algorithms on UTM. Now we can start exhaustive
searching for efficient bounded algorithms of NPC problems by machine
enumeration. The basic idea is that, for input size $n$, first generate
all the possible programs of length less than $UB$, and also generate
all the program input of size upto $n$; then for each program, feed
all the inputs into it, and run the program $prog$ for upto $UB-prog.length$
steps, if it returns all correct answers on those inputs, then add
the program to output list.

Finally we output all the correct $n$-bounded algorithms (sorted
with the smallest $g_{n}^{u(n)}$ value of the worst running time
at first) for further analysis or human inspection, e.g. using machine
aided extrapolation to check if there is any efficient \emph{unbounded}
algorithms.

\subsection{Search by enumeration}

Let us continue to use the 3-coloring problem. The above description
can be formalized by the following algorithm:

\begin{lstlisting}
known_inefficients = {SIMPLE, PRECOMPUTE}
while True:
  UB = minwet(known_inefficients, n)
  outputs = []
  for length in [1, UB]:
    programs = generate_programs_from_strings_with(length)
    for prog in programs:
      results = []
label:input_loop
      for size in [1, n]:
        prog.worst_running_time[size] = 0
        inputs = generate_all_inputs_with(size)
        for input in inputs:
          max_steps = UB - prog.length
          (result, actual_steps) = run(prog, input, max_steps)
          results.append(result, (size,actual_steps))
          if not(result.correct):
            break input_loop
          if actual_steps > prog.worst_running_time[size]:
            prog.worst_running_time[size] = actual_steps
      if all(results correct):
        outputs.append(prog)
  sort(outputs, by(u(n)))
  want_improve = analyze(outputs)
  if want_improve:
    update_knowledge_base(known_inefficients)
  else:
    return outputs
\end{lstlisting}

Line 1, seed the set $known\_inefficients$ with two known inefficient
programs SIMPLE, and PRECOMPUTE.

Line 3, set the initial upper bound $UB$.

Line 5-7, enumerate all the programs with length up to $UB$.

Line 8-13, For each of the program $prog$, feed all the inputs of
size $[1..n]$ one by one.

Line 15, for each input, run the program upto $(UB-prog.length)$
steps.

Line 16, record the pair $(result,(n,step(n)))$.

Line 21-22, if all the returned results are correct, then record $prog$

Line 23-24, output the findings for further analysis, or human inspection.

Line 25-28, if we want to search further, update the program knowledge
base $known\_inefficients$, and continue the next search iteration.

\subsubsection{Update program knowledge base}

The outputs from the previous step may contain bounded programs that
are running time efficient, but loading time inefficient, for example,
there may be a program that has the same running time but half the
length of PRECOMPUTE, and whose length still holds the exponential
relationship with input size $n$. We need to analyze such programs
by human, and if it is found to be inefficient, we add it to the program
knowledge base $known\_inefficients$. This will lower the program
length upper bound $UB=minwet(known\_inefficients,n)$; we will re-run
the enumeration process after such knowledge base update. Fortunately,
as the total number of outputs is finite, this knowledge base updating
process will stop when the knowledge base become saturated.

\subsubsection{Expand search horizon}

Also during the search process, there may be programs that have big
constant loading time, but polynomial running time for $[n_{0},+\infty)$
for $n_{0}>n$, which we will skip. However, this is not a real limitation
of our approach, as we keep increasing $n$ to expand our search horizon,
these programs will be examined again at that time. With the computing
resource increases and implementation improves, the searchable $n$
will keep increasing, and we will gain more knowledge about bounded
algorithms.

\subsection{Machine aided extrapolation}

There are many interesting things can be done to check the output
program's various properties. Probably the most interesting are:
\begin{enumerate}
\item whether any output is actually a potential \emph{unbounded} algorithm,
and
\item whether its running time complexity is \emph{polynomial}.
\end{enumerate}
We can perform these two tests with the help from computer: given
the output of bounded algorithms, we can feed bigger inputs into them,
and check if they will continue return correct results. This can be
formalized by the following algorithm:

\begin{lstlisting}
def analyze(programs, n):
  m = 2 * n
  UB = minwet(known_inefficients, m)
  for prog in programs:
    results = []
label:input_loop
    for size in [n+1, m]:
      inputs = generate_all_inputs_with(size)
      for input in inputs:
        max_steps = UB - prog.length
        result = run(prog, input, max_steps)
        results.append(result)
        if result = undefined:  # i.e. bounded
          break input_loop
    if all(results correct):
      prog.potential_unbounded = true
      if u(m) <= u(n):
        prog.potential_efficient = true
  candidates = filter(programs, by(potential_unbounded))
  return human_inspect(sort(candidates, by(potential_efficient))
\end{lstlisting}

Line 1-11, for each output program $prog$, continue feed all the
problem input size of $[n+1,m]$, (e.g. set $m=2\times n$), and run
it for upto $UB-prog.length$ steps

Line 13-14, if $prog$ return $undefined$, $prog$ is a bounded bounded
algorithm, we can just skip it

Line 15-16, if $prog$ continue return all correct results, then mark
$prog$ as a potential unbounded algorithm

Line 17-21, if the growth rate characteristic function of the running
time of program $prog$ continues to appear to be efficient, then
$prog$ is a potential efficient unbounded algorithm. We give $prog$
high priority for human inspection.

If it is proved to be unbounded efficient, then $prog$ is the program
that can solve NP problem in P time. The author of this paper believes
that we will not find such an algorithm, but would be very happy to
see a pleasant surprise.

\section{Discussions}

\subsection{Does there exist any bounded algorithm with polynomial execution
time for NPC problems?}

This paper has proposed a thought experiment to search for efficient
bounded algorithms of NPC problems by machine enumeration, however
the author is more interested in knowing, and hence would like to
raise the theoretical question:
\begin{problem}
Does there exist any bounded algorithm with polynomial execution time
for NPC problems?
\end{problem}
Although this question is weaker than the original P ?= NP question,
it has the same importance in practice. After all in this real world
we human only have limited resource to build such programs if they
exist.

Compare with the original P ?= NP question, can we take advantage
of the extra program length constraint, and develop some new techniques
to find a proof?

\subsection{Connections to speedup theorems and algorithmic information theory}

In computational complexity theory, the linear speedup theorem for
Turing machines states that for any $TM$ solving a problem with $t(n)$
running time, and for any $c>0$, we can build an equivalent $TM'$
that can solve the same problem with $ct(n)+n+2$ running time. If
we take a closer look of the proof of this theorem (e.g. in \citep{Papadimitriou94}
ch-2.4), and check how the new $TM'$ is constructed, we will see
that the running time speed up is achieved at the expense of increased
machine length (i.e. program length). Blum's speedup theorem \citep{journals/jacm/Blum67}
works the same way by adding ``short-cuts'' to the $TM$'s control
table also using the precompute technique. However running time speedup
does not always mean increased program length. We can achieve both
shorter program length and faster running speed at the same time by
using the precompute-and-cache technique, e.g. finding the number
of 3-colorable graphs with $n$-nodes.

In algorithmic information theory, the Kolmogorov complexity \citep{DBLP:journals/tcs/Kolmogorov98}
of a string (a program in our case) is defined as the length of the
shortest program that can generate the string; while this paper tries
to connect a program's length (the information / knowledge formally
encoded in it) with its running time efficiency.

\subsection{Possible implementation considerations}

In practice we have programs with length of many thousands or millions
of bytes, it will take prohibitively expensive resource to enumerate
them, so the idea proposed in this paper is more a thought experiment.
But if we can start work on small input size, and develop techniques
to reduce the search space. E.g. if we can decide: there is no efficient
bounded program of 3-coloring problem for input size of 4, 5, 6. etc.
Then before we can find a solution to the P/NP problem, we will gain
some knowledge as the exploration length increases. There are many
areas can be developed to speed up and improve the enumeration and
evaluation process, for example:
\begin{itemize}
\item Choose a Universal Turing Machine and a NPC problem with appropriate
properties to reduce the enumeration time or space requirements.
\item Develop techniques to reduce the number of programs we need to search.
For example let us consider all the possible 64-bit long strings,
which correspond to $2^{64}$ Turing Machines (i.e. programs), probably
most of them do not specify valid programs or will abort when being
executed. We can skip generating such invalid program strings from
the beginning.
\item Develop other machine checkable criterion to decide a generated program's
length and time complexity.
\end{itemize}

\section*{Acknowledgment}

The author would like to thank Tony Hoare for his comments on an early
draft of this paper.

\bibliographystyle{plainnat}
\addcontentsline{toc}{section}{\refname}\bibliography{pnp}

\end{document}